\documentclass[hyper]{JHEP} % 10pt is ignored!

\usepackage{epsfig}

\newcommand\fverb{\setbox\pippobox=\hbox\bgroup\verb}

\newcommand\fverbdo{\egroup\medskip\noindent%

            \fbox{\unhbox\pippobox}\ }

\newcommand\fverbit{\egroup\item[\fbox{\unhbox\pippobox}]}

\newbox\pippobox

\title{Supertube Dynamics in Diverse Backgrounds}
\author{J. Kluso\v{n}\\

Department of Theoretical Physics and Astrophysics\\
Faculty of Science, Masaryk University\\
Kotl\'{a}\v{r}sk\'{a} 2, 611 37, Brno, \ \ Czech Republic\\

    E-mail: \email{klu@physics.muni.cz}}
\author{Kamal L. Panigrahi\\

Dip. di Fisica, \& I.N.F.N. Sez. di Roma 2, ``Tor Vergata" \\
Via della Ricerca Scientifica, 1, 00133  Roma, \ \ ITALY\\

    E-mail: \email{Kamal.Panigrahi@roma2.infn.it}}

\preprint{ROM2F/2005/09 \\
\hepth{0506012}} \abstract{We study the homogeneous and time
dependent dynamics of the supertube in diverse backgrounds. After
deriving a general form of the Hamiltonian in general background,
we use a particular gauge fixing, relevant to our analysis to
derive a simpler Hamiltonian. We then study the homogeneous
solutions of the equations of motion in various backgrounds and
study the effective potential in detail.} \keywords{D-branes}

\keywords{D-branes}

\def\bx{\mathbf{x}}

\def\bA{\mathbf{A}}
\def\mE{\mathcal{E}}
\def\mK{\mathcal{K}}
\def\mH{\mathcal{H}}

\begin{document}
%%%%%%%%%%%%%%%%%%%%%
%%%%Introduction %%%%%%%%%
\section{Introduction}\label{first}
The formulation of string theory in various time-dependent and
cosmological backgrounds remains as one of the most important and
exciting open problems for theoretical physicists. In doing so,
one expects a better understanding of some of the outstanding
problems and confusions of quantum gravity and cosmology. In an
attempt along this direction, Sen \cite{Sen:2002nu} proposed a
rolling tachyon solution --the decay of the unstable D-brane
(brane-anti brane pair) when the tachyon rolls in the valley into
the bottom of its potential (vacuum without D-branes). It has also
been argued in \cite{Sen:2002in} that the final state of such a
decay process leads to a classical matter state, with the equation
of motion of a non-interacting and pressureless dust known as
'tachyon mater' with no obvious open string excitations. These
solutions, in general, are constructed by perturbing the boundary
conformal field theory that describes the D-brane by an exact
marginal deformation. The real time tachyon dynamics shows that
the effective field theory for Dirac-Born-Infeld type captures
surprisingly well many aspects of rolling tachyon solutions of
full string theory. (See \cite{Sen:2004nf} and references therein
for a detailed review of the tachyon dynamics)

Recently, Kutasov \cite{Kutasov:2004dj, Kutasov:2004ct} gave a
'geometric realization' of the open string tachyon, in the form of
the rolling D-brane--the time dependent dynamics of the D-brane in
the throat of NS5-branes. It has been shown that the decay process
resembles astonishingly, that of the decay of unstable D-branes,
when we restricts ourselves to the case when the distance between
the D-brane and NS5-brane is of the order of the string length
$(l_s)$. One wonders whether the above identification is an
artifact of low energy effective field theory or it can be
extended to the full string theory. This was analyzed in
\cite{Nakayama:2004yx} in the form of the boundary states of the
in falling D-branes into the throat region of a stack of
NS5-branes (the CHS geometry). The electric deformation of the
process has also been realized in terms of the decay of the
electrified D-brane into the NS5
throat\cite{Chen:2004vw,Nakayama:2004ge}. This has further been
extended in \cite{Panigrahi:2004qr, Saremi:2004yd} to the case of
D$p$-brane background. Later on in
\cite{Kluson:2004xc,Kluson:2004yk}, this ideas were extended to
the case of the Non-BPS probe branes and a careful analysis
revealed the existence of a symmetry (possibly broken explicitly
at the Lagrangian level \cite{Kluson:2005qx,Kluson:2005jr}). Hence
a natural guess will be that the dynamics of the D-branes in the
curved backgrounds has a much more richer structure in the
geometry and in the dynamics\cite{Kluson:2005dr}, and hence likely
to add some input in our understanding of string theory in time
dependent curved backgrounds. See (\cite{Yavartanoo:2004wb} -
\cite{Thomas:2005fu}) for related studies of the D-brane dynamics
and cosmological applications in various backgrounds.

In another context, the supertube\cite{Mateos:2001qs}--the bunch
of straight strings with D0-brane blown up into a supersymmetric
tubular D2-brane with electric and magnetic field-- has been
instrumental in understanding black hole physics. For example, the
quantum states of a supertube counted from the direct quantization
of the Born-Infeld action near the geometrically allowed
microstates with some fixed charges are shown to be in one to one
correspondence with some black holes. The stability of this bound
state is achieved by the non-zero angular momentum generated by
the Born-Infeld electric and magnetic charges. Recently the
stability of the supertubes in other curved backgrounds has also
been studied in \cite{Huang:2005rd}. Hence in the recent surge of
studying the D-brane dynamics in various backgrounds, in
connection with finding out the time dependent solutions of string
theory and tachyon dynamics, it seems very interesting to study
the dynamics of the supertube in diverse backgrounds. But looking
at the rather complicated analysis of the problem in the usual
effective theory approach, we would like to take advantage of the
Hamiltonian formulation. This approach has been very instructive
in investigating the D-branes in the strong coupling limit, by
formally taking the zero tension
limit\cite{Lindstrom:1997uj,Gustafsson:1998ej}, for example. By
doing this we achieve a formal Hamiltonian for the D-brane motion
in curved background, and the beauty of this formalism makes us
comfortable to use the constraints of equations of motion in
appropriate ways. We try to be as general as possible in the
beginning, but while studying the motion of the tube in the
background of various macroscopic objects, we use some properties
of the supergravity backgrounds relevant for studying string
theory, to make the analysis simpler.

The layout of the paper is as follows. In section-2, we start by
deriving the Dirac-Born-Infeld action using the Hamiltonian
formulation. The rather complicated action can be made simpler by
making use of an appropriate gauge fixing. After deriving an
action for the dynamics of the tube in rather general curved
backgrounds, in section-3, we focus our attention to the case of
Dp-branes, NS5-branes and the fundamental string backgrounds. We
study in detail the effective potential and the motion of the tube
in the vicinity of these backgrounds generated by the macroscopic
objects. We present our conclusions in section-4.
%%%%%%%%%%%%%%%%%%%%%%%%%%%%%%%%%%%%%%%%%%%%%%%%%%%%%%%%%%%%%%%%%%%
%%%%%%%%%%%%%%%%%%%%%%%%%%%%%%%%%%%%%%%%%%%%%%%%%%%%%%%%%%%%%%%%%%%%
\section{Hamiltonian dynamics for
Dp-brane in general background and its gauge fixed version}

This section is devoted to the formulation of the Hamiltonian
formalism for Dp-brane in general curved background. We will
mainly follow the very nice analysis presented in
\cite{Lindstrom:1997uj,Gustafsson:1998ej}. The Dirac-Born-Infeld
action for a Dp-brane in a general bosonic background is given by
the following usual form\footnote{We do not consider the
Wess-Zummino term since in the examples studied below the coupling
to Ramond-Ramond fields is not important.}

\begin{equation}\label{actL}
S_p=-\tau_p\int d^{p+1}\xi e^{-\Phi} \sqrt{-\det\bA} \ ,
\end{equation}
where
\begin{equation}
\bA_{\mu\nu}=\gamma_{\mu\nu}+ F_{\mu\nu}, \>\>\> \mu,\nu=0,\dots,p
\ ,
\end{equation}
and where $\tau_p$ is Dp-brane tension.
The induced metric $\gamma_{\mu\nu}$ and the induced field
strengths on the worldvolume $F_{\mu\nu}$ are given by the
following expressions
\begin{eqnarray}
\gamma_{\mu\nu}&=& \partial_\mu X^M
\partial_\nu X^N g_{MN}, \cr & \cr
F_{\mu\nu}&=& \partial_\mu X^M\partial_\nu X^N b_{MN}
+\partial_\mu A_\nu - \partial_\nu A_\mu \ .
\end{eqnarray}
Where $g_{MN}(X(\xi)) \ , b_{MN}(X(\xi)) \ , M,N=0,\dots,9$
 are metric and NS-NS two
form field that are in general functions of the embedding
coordinates $X^M(\xi)$. Let us now calculate the conjugate momenta
from (\ref{actL})
\begin{eqnarray}
P_M(\xi)&=& \frac{\delta S}{\delta \partial_0 X^M(\xi)}
=-\frac{e^{-\Phi}} {2}\sqrt{-\det\bA}\partial_\nu X^N
\left(g_{MN}(\bA^{-1})^{(\nu 0)} +b_{MN}(\bA^{-1})^{[\nu
0]}\right) \ , \cr & \cr \pi^a(\xi)&=&\frac{\delta S}{\delta
\partial_0 A_a(\xi)}
=-\frac{e^{-\Phi}}{2}\sqrt{-\det\bA} (\bA^{-1})^{[a0]} \ , a, b =
1, \dots, p \ , \cr & \cr \pi^0(\xi)&=& \frac{\delta S}{\delta
\partial_0 A_0(\xi)}=
-\frac{e^{-\Phi}}{2} \left((\bA^{-1})^{00}- (\bA^{-1})^{00}\right)
\sqrt{-\det\bA}=0 \ ,
\nonumber \\
\end{eqnarray}
where we have introduced the symmetric and antisymmetric form of
any (p+1)$\times$(p+1) matrix ${\rm A}$ as
\begin{equation}
{\rm A}^{(\mu\nu)}={\rm A}^{\mu\nu}+{\rm A}^{\nu\mu} \ , {\rm
A}^{[\mu\nu]}={\rm A}^{\mu\nu}-{\rm A}^{\nu\mu} \ .
\end{equation}
For our purpose it is useful to define the conjugate momenta
$\Pi_M$ as
\begin{equation}
\Pi_M=P_M+P^aB_{aM}= -\frac{e^{-\Phi} }{2} V\sqrt{-\det
\bA}\gamma_{M\nu} (\bA^{-1})^{(\nu 0)} \ ,
\end{equation}
where
\begin{equation}
\gamma_{M\nu}=g_{MN}\partial X^N_\nu \ , B_{aM}=\partial_a
X^NB_{NM} \ .
\end{equation}
 Using these expressions it is easy to see
that following constraints are obeyed
\begin{eqnarray}
\Pi_M\partial_a X^M+ \pi^aF_{ab}=0 \ ,
\nonumber \\
\Pi_Mg^{MN}\Pi_N+ \pi^a\gamma_{ab} \pi^b+e^{-2\Phi}\tau_p^2
\det\bA_{ab}=0 \ ,
\nonumber \\
\pi^0=0 \ . \nonumber \\
\end{eqnarray}
The corresponding Hamiltonian for Dp-brane in curved background
takes the following form
\begin{eqnarray}
H=\int d^p\xi \mathcal{H}(\xi) \ ,
\end{eqnarray}
where the
Hamiltonian density $(\mathcal{H})$ is given by
\begin{eqnarray}
\mathcal{H} &=& \pi^i\partial_i A_0 +\sigma \pi^0 +\rho^a
(\Pi_M\partial_a X^M + F_{ab}\pi^b) \cr & \cr &+& \lambda
(\Pi_Mg^{MN} \Pi_N+\pi^a\gamma_{ab}\pi^b
+e^{-2\Phi}\tau_p^2\det\bA_{ab})\ ,
\end{eqnarray}
where $\sigma, \rho^a$ and $\lambda$ are Lagrange multipliers for
the constraints. More precisely, the final Hamiltonian is just the
sum of constraints, in agreement with the diffeomorphism
invariance of the original Lagrangian.
%%%%%%%%%%%%%%%%%%%%%%%%%%%%%%%%%%%%%%%%%%%%%%%%%%%%%%%%%%%%%%%%
\subsection{Gauge fixing problem}
Since the original DBI action is diffeomorphism invariant, it is
convenient to use this  symmetry to reduce the number of
independent equations of motions. This procedure is commonly known
as gauge fixing.

To proceed we can without loss of generality presume that the
metric takes  diagonal form. Then we consider the equation of
motion for $X^M$
\begin{eqnarray}
\partial_0 X^M=
\frac{\delta H}{\delta P_M(\xi)}= \rho^a\partial_aX^M+ 2\lambda
g^{MN}\Pi_N \ .
\nonumber \\
\end{eqnarray}
Usually the static gauge is imposed at the level of the action and
Lagrangian. In our case, however, we would like to perform the
similar procedure at the level of canonical equations of motion.
Since in the Lagrangian formalism the static gauge is imposed by
the constraints $X^{\mu}=\xi^{\mu} \ , \mu=0,\dots,p$,  it is
natural to solve the canonical equations of motion for $X^\mu$ by
the ansatz
\begin{equation}
\partial_0 X^0=1 \ ,
\partial_a X^b=\delta_a^b \ ,
 a=1,\dots,p \ .
\end{equation}
Then the equations of motion for $X^\mu $ simplify as
\begin{eqnarray}
\Pi_0=\frac{g_{00}}{2\lambda} \ , \Pi_a=-\frac{\rho^a g_{aa}}
{2\lambda} \ \ (\mathrm{No \ summation \ over \ a})
 \ .
\nonumber \\
\end{eqnarray}
Using these results we get
\begin{equation}
\Pi_M\partial_a X^M = -\frac{\rho^bg_{ba}}{2\lambda}+
\Pi_s\partial_a X^s
\end{equation}
where $s,r,t\dots$ label the directions transverse to Dp-brane. If
we ignore the terms in the Hamiltonian that enforce Gauss law
constraints and the vanishing of $\pi^0$ we obtain \footnote{To
see this, note that the equation of motion for $A_0$ leads to the
``Gauss law'' constraints
\begin{equation}
\frac{\delta H}{\delta A_0(\bx)}=0 \Rightarrow
\partial_a \pi^a=0 \
\end{equation} and variation of the
Hamiltonian with respect to $\sigma$ implies $\pi^0=0$.}
\begin{eqnarray}\label{hdep}
\mathcal{H}= \frac{g_{00}}{4\lambda} -\frac{\rho^a
g_{ab}\rho^b}{4\lambda} +\rho^a b_a+\lambda D \ , \nonumber \\
\end{eqnarray}
where
\begin{eqnarray}
b_a=\Pi_s\partial_a X^s  +F_{ab}\pi^b
\ , \nonumber \\
D=\Pi_rg^{rs}\Pi_s+ \pi^a\gamma_{ab}\pi^b+e^{-2\Phi}
\tau_p^2\det\bA_{ab} \ .
\nonumber \\
\end{eqnarray}
In order to enforce the variation with the Lagrange multiplies we
should consider the Dp-brane action
\begin{eqnarray}
S_p=-\int d^{p+1}\xi \> \mathcal{L}
\end{eqnarray}
where we express $\mathcal{L}$ as a Lagrange transformation of the
hamiltonian density (\ref{hdep}) so that we obtain the action in
the form
\begin{equation}
S_p=-\int  d^{p+1}\xi \left(P_M\partial_0 X^M+ \pi_a\partial_0
A^a-\mathcal{H} \right)
\end{equation}
Now we replace $P_\mu$ with the help of $\Pi_\mu$ given above.
Using the fact that $\partial_0 X^\mu= \delta_0^\mu$ we obtain an
action for $P_s,X^s$ that contains the Lagrange multipliers
$\lambda,\rho^a$
\begin{eqnarray}
S_p = -\int d^{p+1}\xi \left(P_s\partial_0 X^s+ \pi^a\partial_0A_a
-\pi^aB_{a0}+ \frac{g_{00}}{4\lambda}
+\frac{\rho^ag_{ab}\rho^b}{4\lambda} -\lambda D -\rho^ab_a\right).
\nonumber \\
\end{eqnarray}
Now the variation with respect to $\rho^a$ implies (
remember that $g_{ab}$ is diagonal)
\begin{equation}\label{rhog}
\rho^a=2\lambda g^{ab} b_b \ .
\end{equation}
If we then insert (\ref{rhog}) back into the action we get
\begin{equation}
S=-\int d^{p+1}\xi \left(P_s\partial_0 X^s
+\pi^a\partial_0A_a-\pi^aB_{a0} +\frac{g_{00}}{4\lambda}
-\lambda(D+b_ag^{ab}b_b)\right) \ .
\end{equation}
Now the variation with respect to $\lambda$ implies
\begin{equation}
\lambda=\frac{1}{2}\sqrt{\frac{-g_{00}} {b_ag^{ab}b_b+D}} \ .
\end{equation}
Inserting back to the action we get finally
\begin{equation}
S=-\int d^{p+1}\xi \left(P_s\partial_0 X^s +\pi^a
\partial_0 A_a
-\sqrt{-g_{00}} \sqrt{b_ag^{ab}b_b+D}-\pi^aB_{a0} \right)
\end{equation}
from which we obtain the Hamiltonian density for transverse
variables $Y^s,P_s$ and
  gauge field $p^a,A_a$ in
the form
\begin{eqnarray}\label{Hdeng1}
\mH &=& \sqrt{-g_{00}} \sqrt{\mK}+\Pi^aB_{a0} \ , \cr & \cr \mK&=&
\Pi_rg^{rs}\Pi_s+ \pi^a\gamma_{ab}\pi^b+
b_ag^{ab}b_b+e^{-2\Phi}\tau_p^2 \det\bA_{ab} \  , \cr & \cr \Pi_s
&=& P_s+\pi^aB_{as} \ , b_a=\Pi_s\partial_aX^s+ F_{ab}\pi^b \ .
\end{eqnarray}

Now we are ready, using the Hamiltonian density (\ref{Hdeng1}) to
study the dynamics of the supertube in the various supergravity
background.
%%%%%%%%%%%%%%%%%%%%%%%%%%%%%%%%
%%%%%%%%%%%%%%%%%%%%%%%%%%%%%%%%%%%%%%%%%%%%%%%%%%%%%%%%%%%%%%%%
\section{Supertube in Diverse
Backgrounds} By standard treatment, the supertube in flat
spacetime is the D2-brane that is stretched in one particular
direction, say $z$ direction and that have arbitrary shape in the
transverse $R^8$ space. In the case when we study the supertube
dynamics in the background of some macroscopic objects (Dp-branes,
NS5-branes or fundamental strings) the situation is slightly more
complicated. On the other hand since
 these backgrounds have generally  manifest
rotational invariance of the transverse $R^{9-k}$ space, where $k$
is the spatial dimension of given objects,   it is natural to
simplify the analysis by presuming that the supertube has its base
in the $(x^8,x^9)$-plane \footnote{ We  restrict ourselves
 to the case of background Dp-branes with
 $p<5$ so that the coupling of the
D2-brane to the background   Ramond-Ramond field vanishes.}. In
this plane we introduce the polar coordinates
\begin{equation}
x^8= R \cos\phi \  , x^9= R \sin\phi \ .
\end{equation}
Then the embedding coordinates are $R(\sigma,t) \ , X^m \ ,
m=p+1,\dots,7$, that parameterize the position of D2-brane in the
directions transverse to given macroscopic objects of spatial
dimension $p$ which are however transverse to the $(x^8,x^9)$
plane. We also introduce the coordinates $ Y^u \ ,
(u,v=2,\dots,p)$, where $Y^u$ parameterize the position of
supertube in the space parallel with Dp-brane. Since the $B$ field
is also zero, the  Hamiltonian for such a Dp-brane takes the form
\begin{equation}\label{Hstu}
H=\int dt dz d\sigma \mathcal{H} = \int dt dz d\sigma
\left[\sqrt{-g_{_{00}}}\sqrt{\mK}\right] \ ,
\end{equation}
with
\begin{eqnarray}
\mK &=& \pi_a g^{ab}\pi_b+
 p_{_R} g^{RR} p_{_R}+
p_mg^{mn}p_n +p_ug^{uv}p_v + b_ag^{ab}b_b \cr & \cr &+&
(\pi^a\partial_aR) g_{_{RR}}(\pi^b
\partial_bR)+
(\pi^a\partial_aY^u)g_{_{uv}} (\pi^b\partial_b Y^v) \cr & \cr
&+&(\pi^a\partial_a X^m)g_{mn} (\pi^b\partial_b X^n)+ e^{-2\Phi}
\tau_2^2\det \bA_{ab}  \ ,
\end{eqnarray}
and
\begin{equation}
\bA_{ab}=g_{_{ab}}+g_{{_R} {_R}}\
\partial_aR\partial_bR+g_{mn}
\partial_a X^m\partial_b X^n
+g_{_{uv}}\partial_aY^u\partial_bY^v + F_{_{ab}} \ , a,b=z,\phi
\end{equation}
with
\begin{equation}
b_a=F_{_{ab}}  \pi^b+
\partial_a  R \ p_{_R} +
\partial_a X^mp_m+
\partial_a  Y^u
p_{_u} \ .
\end{equation}
Using the Hamiltonian  (\ref{Hstu}) the equation of motions for
$A_a$ and $\pi^a$ take the form
\begin{equation}
\dot{A}_a(\bx)=\frac{\delta H} {\delta \pi^a(\bx)} =\sqrt{-g_{00}}
\frac{g_{ab}\pi^b+\partial_aX^m g_{mn}(\pi^b\partial_bX^n)
+\partial_a Y^ug_{uv}(\pi^b\partial_b Y^v) +F_{ab} g^{bc}b_c}
{\sqrt{\mK}} \ ,
\end{equation}
\begin{eqnarray}
\dot{\pi}^a(\bx)&=& -\frac{\delta H} {\delta A_a(\bx)} =
\partial_c\left[\frac{\sqrt{-g_{00}}
\pi^ag^{cb}b_b}{\sqrt{\mK}}\right] -\partial_c\left[
\frac{\sqrt{-g_{00}} \pi^cg^{ab}b_b}{\sqrt{\mK}}\right] \cr & \cr
&+&\frac{1}{2}\partial_c\left[\frac{e^{-2\Phi}
\tau_2^2\sqrt{-g_{00}}}{\sqrt{\mK}} (\bA^{-1})^{ac}\det \bA_{ab}\right]
-\frac{1}{2}\partial_c\left[\frac{e^{-2\Phi}
\tau_2^2\sqrt{-g_{00}}}{\sqrt{\mK}} (\bA^{-1})^{ca}\det \bA_{ab}\right]
\ .
\nonumber \\
\end{eqnarray}
Further, the equations of motion for the embedding coordinates are
\begin{eqnarray}
\dot{R}(\bx)&=& \frac{\delta H}{\delta p_R(\bx)}=
\frac{\sqrt{-g_{00}}}{\sqrt{\mK}}
\left(g^{RR}p_R+\partial_iRg^{ab}b_b \right) \ , \cr & \cr
\dot{X}^m(\bx) &=& \frac{\delta H}{\delta p_m(\bx)}=
\frac{\sqrt{-g_{00}}}{\sqrt{\mK}} \left(g^{mn}p_n+
\partial_aX^mg^{ab}b_b
\right) \ , \cr & \cr \dot{Y}^u(\bx) &=& \frac{\delta H}{\delta
p_u(\bx)}= \frac{\sqrt{-g_{00}}}{\sqrt{\mK}} \left(g^{uv}p_v+
\partial_aX^ug^{ab}b_b
\right),
\end{eqnarray}
while the equation of motion for $p_m\ , p_u$ and $p_R$ are
{\small
\begin{eqnarray}
&&\dot{p}_m(\bx) = -\frac{\delta H}{\delta X^m(\bx)}=
\partial_a\left[
\frac{\sqrt{-g_{00}}}{\sqrt{\mK}}
\pi^ag_{mn}(\pi^b\partial_bX^n)\right] \cr & \cr
&+&\partial_a\left[\frac{\sqrt{-g_{00}}}
{\sqrt{\mK}}e^{-2\Phi}\tau_2^2g_{mn}
\partial_b X^n(\bA^{-1})^{ba}\det\bA_{ab}
\right] + \frac{\delta g_{_{00}}}{\delta X^m}
\frac{\sqrt{\mK}}{2\sqrt{-g_{_{00}}}} \cr & \cr &-&
\frac{\sqrt{-g_{_{00}}}}{2\sqrt{\mK}} \Big(\pi^a\frac{\delta
g_{ab}}{\delta X^m} \pi^b+p_R\frac{\delta g^{RR}}{\delta X^m}p_R
+p_n\frac{\delta g^{no}}{\delta X^m}p_o +p_u\frac{\delta
g^{uv}}{\delta X^m}p_v +\frac{\delta [e^{-2\Phi}]}{\delta
R}\tau_2^2 \det\bA_{ab} \cr & \cr &+&
e^{-2\Phi}\tau_2^2\left[\frac{\delta g_{_{RR}}}{\delta X^m}
(\partial_aR\partial_b R) + \frac{\delta g_{no}}{\delta X^m}
(\partial_a X^n\partial_bX^o) + \frac{\delta g_{uv}}{\delta X^m}
(\partial_a Y^u\partial_bY^v) \right] (\bA^{-1})^{ba}
\det \bA_{ab} \Big)  \ ,  \nonumber \\
\end{eqnarray}}
\begin{eqnarray}
\dot{p}_u(\bx) = -\frac{\delta H}{\delta Y^u(\bx)}&=&
\partial_a\left[
\frac{\sqrt{-g_{00}}}{\sqrt{\mK}}
\pi^ag_{uv}(\pi^b\partial_bY^v)\right]\cr & \cr &+&
\partial_a\left[\frac{\sqrt{-g_{00}}}
{\sqrt{\mK}}e^{-2\Phi}\tau_2^2g_{uv}
\partial_b Y^v(\bA^{-1})^{ba}\det\bA_{ab}
\right] \nonumber \\
\end{eqnarray}
and {\small
\begin{eqnarray}
&&\dot{p}_{_R}(\bx)= -\frac{\delta H}{\delta R(\bx)}
=\partial_a\left[\frac{\sqrt{-g_{_{00}}}}{\sqrt{\mK}}
\pi^ag_{_{RR}}(\pi^b\partial_bR)\right]\cr & \cr &+&
\partial_a\left[\frac{\sqrt{-g_{_{00}}}}
{\sqrt{\mK}}e^{-2\Phi}\tau_2^2g_{_{RR}}
\partial_b R(\bA^{-1})^{ba}\det\bA_{ab}
\right]+ \frac{\delta g_{_{00}}}{\delta R}
\frac{\sqrt{\mK}}{2\sqrt{-g_{_{00}}}}\cr & \cr
&-&\frac{\sqrt{-g_{_{00}}}}{2\sqrt{\mK}} \Big(\pi^a\frac{\delta
g_{ab}}{\delta R} \pi^b+p_R\frac{\delta g^{RR}}{\delta R}p_R
+p_m\frac{\delta g^{mn}}{\delta R}p_n +p_u\frac{\delta
g^{uv}}{\delta R}p_v +\frac{\delta [e^{-2\Phi}]}{\delta R}\tau_2^2
\det \bA_{ab}\cr & \cr &+& e^{-2\Phi}\tau_2^2\left[\frac{\delta
g_{_{RR}}}{\delta R} (\partial_aR\partial_b R) + \frac{\delta
g_{mn}}{\delta R} (\partial_a X^m\partial_bX^n) + \frac{\delta
g_{uv}}{\delta R} (\partial_a Y^u\partial_bY^v) \right]
(\bA^{-1})^{ba}
\det \bA_{ab} \Big) \nonumber \\
\end{eqnarray}}
Note that generally the metric is function of the expression
$R^2+X^mX_m$. On the other hand thanks to the symmetry of the
problem with respect to $z$ direction it is natural to consider
the modes that are $z$ independent. We also take  the ansatz for
the gauge field in the form
\begin{equation}
F=\partial_0A_z(\phi) dt\wedge dz+ B(\phi)dz\wedge d\phi \ ,
A_\phi=Bz \ .
\end{equation}
Then the matrix  $\bA_{ab}$ takes the form
\begin{equation}
\bA_{ab}=\left(\begin{array}{cccc}
g_{_{zz}} & & & B \\
-B & & & g_{_{\phi\phi}}+g_{_{RR}} (\partial_\phi R)^2+
g_{mn}\partial_\phi X^m\partial_\phi X^n+ g_{_{uv}}\partial_\phi
Y^u\partial_\phi Y^v\\ \end{array} \right)
\end{equation}
so that
\begin{equation}
\det \bA_{ab}= g_{_{zz}}\left(g_{_{\phi\phi}}+
g_{_{RR}}(\partial_\phi R)^2+ g_{mn}\partial_\phi X^m\partial_\phi
X^n+ g_{_{uv}}\partial_\phi Y^u\partial_\phi Y^v\right) + B^2
\end{equation}
and
\begin{equation}
(\bA^{-1})^{ab} = \frac{1}{\det \bA} \left(
\begin{array}{cccc}
 g_{_{\phi\phi}}+
g_{_{RR}}(\partial_\phi R)^2+ g_{mn}\partial_\phi X^m\partial_\phi
X^n+
 g_{_{uv}}\partial_\phi
Y^u\partial_\phi
Y^v & & & B \\
-B & & & g_{_{zz}} \\ \end{array}\right) \ .
\end{equation}
 With this notation the equation of motion for $\pi^z$
takes the form
\begin{equation}\label{dotpz}
\dot{\pi}^z=
\partial_\phi\left[
\frac{\sqrt{-g_{_{00}}}}{\sqrt{\mK}}
g^{\phi\phi}(-B\pi^z+\partial_\phi X^mp_{_m} +\partial_\phi Y^up_u
) \right]+
\partial_{_\phi}\left[e^{-2\Phi}\tau_2^2
\frac{\sqrt{-g_{_{00}}}B}{\sqrt{\mK}}\right]
\end{equation}
As it is clear from the equation above  $\pi^z$ is generally time
dependent. On the other hand we can show that the quantity
$n=\oint d\phi \pi^z$ is conserved since
\begin{equation}
\dot{n}=\oint d\phi \dot{\pi}^z= \oint d\phi \partial_\phi[\dots]
=0 \ .
\end{equation}
The similar case occurs for $p_u$ where
 $\dot{p}_u(\bx)\neq 0$
\begin{equation}\label{dotpu}
\dot{p}_u=\partial_\phi \left[\frac{\sqrt{-g_{00}}}{\sqrt{\mK}}
e^{-2\Phi}\tau_2^2g_{uv}(\partial_\phi Y^v) g_{zz}\right]
\end{equation}
while the  the total momentum $P_u$ is conserved
\begin{equation}
P_u\equiv \oint d\phi p_u \Rightarrow \dot{P}_u= \oint
d\phi\dot{p}_u= \oint d\phi \partial_\phi[(\dots)]=0
\end{equation}
To simplify further the analysis of the time dependent evolution
of supertube we will restrict ourselves to the case of homogenous
fields on the worldvolume of D2-brane:
\begin{equation}
\partial_\phi R=\partial_\phi X^m=\partial_\phi Y^u=
\partial_\phi \pi=0
\end{equation}
As it is now clear
 from (\ref{dotpu}) and (\ref{dotpz})
  $\pi^z\equiv \Pi$ and $p_u$ are conserved. On the other hand
the equation of motion for $A_z \ , A_\phi$ are equal to
\begin{eqnarray}
\dot{A}_z= E=\frac{\sqrt{-g_{00}}}{\sqrt{\mK}}
(g_{zz}+g^{\phi\phi}B^2)\Pi^2 \ ,
\nonumber \\
\dot{A}_\phi=0 \ , \nonumber \\
\end{eqnarray}
where the second equation above implies that $B=\partial_zA_\phi$
is conserved as well. Then we get simple form of the Hamiltonian
density $\mH$
\begin{equation}
\small{ \mH=\sqrt{-g_{00}} \sqrt{p_Rg^{RR}p_R+ p_mg^{mn}p_n+
p_ug^{uv}p_v +\Pi g_{zz}\Pi+ b_\phi g^{\phi\phi}b_\phi+
e^{-2\Phi}\tau_2^2 \left(g_{zz}g_{\phi\phi}+B^2\right)}} \ ,
\end{equation}
where
\begin{equation}
b_\phi=F_{\phi z}\pi^z= -B\Pi
\end{equation}
For supertube in D$p$-brane background we have
\begin{eqnarray}
g_{00} &=& -H_p^{-1/2} \ , g_{zz}=H_p^{-1/2} \ ,
g_{\phi\phi}=H_p^{1/2}R^2 \ , \cr & \cr g_{uv}&=&
H_p^{-1/2}\delta_{uv} \ , g_{mn}=H_p^{1/2}\delta_{mn} \ ,
g_{RR}=H_p^{1/2} \ , \cr & \cr e^{-2\Phi} &=& H_p^{\frac{p-3}{2}}
\end{eqnarray}
where the harmonic function for $N$ D$p$-branes is given by
\footnote{We work in units $l_s=1$.}
\begin{equation}
H_p=1+\frac{N}{(R^2+X^mX_m)^{(7-p)/2}}
\end{equation}
and hence the Hamiltonian density takes the form
\begin{equation}
\mH=\sqrt{\frac{\Pi^2}{H_p}
\left(1+\frac{B^2}{R^2}\right)+\frac{p_R^2}{H_p}
+\frac{p_mp^m}{H_p}+ p_up^u+\tau_2^2H_p^{\frac{p-4}{2}}
\left[R^2+B^2\right]}
\end{equation}
that in the static case ($p_u=p_R=p_m=0$) reduces to the
Hamiltonian studied in \cite{Huang:2005rd}. For letter purposes it
is also useful to define conserved energy $\mE$ through the
relation
\begin{equation}
\mE= \oint \mH=2\pi \mH
 \Rightarrow \mH=\frac{\mE}{2\pi} \ .
\end{equation}
As it is clear from the form of the harmonic function $H_p$ the
background has $SO(7-p)$ symmetry in the transverse subspace
labelled with coordinates $X^m$. Then it is natural to simplify
the analysis by restricting the dynamics of the supertube to the
$(x^6,x^7)$ plane and introduce second polar coordinates as
\begin{equation}
x^6=\rho \cos \psi \ , x^7=\rho \sin \psi \
\end{equation}
with corresponding metric components
\begin{equation}
g_{\rho\rho}=H_p^{1/2} \ , g_{\psi\psi}=H_p^{1/2}\rho^2 \ .
\end{equation}
Then the Hamiltonian  density $\mH$  can be written as
\begin{equation}
\mH=\sqrt{\frac{\Pi^2}{H_p}
\left(1+\frac{B^2}{R^2}\right)+\frac{p_R^2}{H_p}
+\frac{p_\rho^2}{H_p}+ \frac{p_\psi^2}{H_p\rho^2}+
p_up^u+\tau_2^2H_p^{\frac{p-4}{2}} \left[R^2+B^2\right]} \ .
\end{equation}
Since the Hamiltonian does not explicitly depend on $\psi$ we get
immediately that $p_\psi$ is conserved:
\begin{equation}
\dot{p}_\psi=-\frac{\delta H}{\delta \psi}=0 \ .
\end{equation}
Then the  equations of motion for $\rho,R$ take the form
\begin{eqnarray}
\dot{R} &=&\frac{\delta H}{\delta p_R}= \frac{2\pi p_R} {H_p\mE} \
, \cr & \cr \dot{\rho}&=& \frac{\delta H}{\delta p_\rho}=
 \frac{2\pi p_\rho}
{H_p\mE} \  \nonumber \\
\end{eqnarray}
On the other hand the equation of motion for $p_{_R},p_{_\rho}$
are much more complicated thanks to the nontrivial dependence of
$H$ on $\rho,R$. In fact it is very complicated to solve these
equations in the full generality. Let us rather consider the
special case when we will study the dynamics of $R$ only. In order
to do this we should find the stable values for $\rho$. Let us
then consider the equation of motion for $p_\rho$
\begin{eqnarray}
\dot{p}_\rho = -\frac{1}{2\sqrt{\mK}} \frac{\delta \mK} {\delta
H_p}\frac{\delta H_p}{\delta \rho} +\frac{p^2_\psi}{\rho^3H_p
\sqrt{\mK}}= -\frac{1}{\sqrt{\mK}} \left[\frac{N(p-7)\rho}
{(R^2+\rho^2)^{(8-p)/2}}\frac{\delta \mK}{\delta
H_p}-\frac{p^2_\psi}{\rho^3
H_p}\right] \nonumber \\
\end{eqnarray}
We see that the momentum $p_\rho$ is equal to zero for
$\rho=\psi=0$. The question is whether there exist closed orbits
with $p_\psi\neq 0$ for which $p_\rho=0$. Looking at the equation
above it is clear that we find $\rho$ as function of $N$,
conserved momenta $p_u,p_\psi$ and, most importanly, as a function
of $R$. Then it follows that the resulting Hamiltonian is
complicated function of $R$. Even if it would be certainly
interesting to study the properties of supertube with nonzero
$p_\psi$ we will restrict ourselved in this paper to the case of
$p_\psi=\rho=0$.

Analogously, we can consider the situation when $p_{_R}=0$ that
corresponds some particular value of
$R_{stat}=R(N,p_u,p_\psi,\rho)$ and study the time dependence of
$\rho$. As in the previous case we leave the study of this problem
for future.

Now let us consider the case when $p_\rho=p_\psi= \rho=0$. In this
case the study of the supertube dynamics reduces to the study of
the time dependence of $R$. Then it is natural to write the
Hamiltonian density as as
$\mH=\sqrt{\frac{p_R^2}{H_p} +V}$ where
$\sqrt{V}=\mH(p_{_R}=0)$. Then
 we can write the equation of motion for
$R$ in the following form
\begin{equation}
\dot{R}=\frac{1}{\mE \sqrt{H_p}} \sqrt {\mE^2-(2\pi)^2V} \ ,
\frac{2}{2}\dot{R}^2+V_{\rm{eff}}=0 \ ,
\end{equation}
where
\begin{equation}
V_{\rm{eff}}=\frac{1}{H_p} \left( \frac{(2\pi)^2V} {\mE^2}- 1
\right) \ .
\end{equation}
Then in order to obtain qualitative character of the dynamics of
he supertube we use the observation \cite{Burgess:2003mm} that the
equation given above corresponds to the conservation of energy for
massive particle with mass ($m=2$) with the effective potential
$V_{\rm{eff}}$ with total zero energy. We will be interested in
two cases corresponding to $p=2,4$.
\subsection{D2-brane background}
In this case the effective potential takes the form
\begin{equation}
V=(R^2+B^2)\left(\frac{R^3}{ R^5+N}\right)
 (\Pi^2+\tau_2^2R^2)+p_up^u
\end{equation}
and hence $V_{\rm{eff}}$ is equal to
\begin{equation}
V_{\rm{eff}}=\frac{R^5} {R^5+N} \left(\frac{4\pi^2} {\mE^2}
(R^2+B^2)\frac{R^3}{ R^5+N} (\Pi^2+
\tau_2^2R^2)+\frac{4\pi^2p_up^u} {\mE^2}- 1\right)
 \ .
\end{equation}
It is clear that the particle with zero energy can move in the
interval between the points where $V_{\rm{eff}}=0$. First of all,
the asymptotic behavior of $V_{\rm{eff}}$ is as follows
\begin{eqnarray}
V_{\rm{eff}}\approx \frac{R^5}{N} (\frac{4\pi^2B^2 \Pi^2
\tau_2R^3}{\mE^2}+\frac{4\pi^2 p_up^u} {\mE^2}-1) \ ,
R\rightarrow 0 \ , \nonumber \\
V_{\rm{eff}}\approx \frac{4\pi^2}{\mE^2}R^2
\ , R\rightarrow \infty \ . \nonumber \\
\end{eqnarray}
Few comments are in order. From the above limits we can see that
the potential approaches $0$ for $R\rightarrow 0$ from below. Then
since the potential blows up for $R\rightarrow \infty$ there
should exists the point $R_T$ where the potential vanishes:
$V_{\rm{eff}}(R_T)=0$. In order to study the dynamics around this
point we  introduce the variable $r$ through the substitution
$R=r+R_T$ and insert it to the expression for conservation of
energy. Using the fact that near the  turning point we have \\
$V_{\rm{eff}}(R)=V_{\rm{eff}}(R_T)+V'_{\rm eff}(R_T)r =V'_{\rm
eff}(R_T)r$ where we have used $V_{\rm{eff}}(R_T)=0$. Then we get
the equation of motion for $r$
\begin{equation}
\dot{r}^2=-V'_{\rm{eff}}(R_T)r \Rightarrow
\dot{r}=\pm\sqrt{-V_{\rm{eff}}'(R_T)r} \ .
\end{equation}
Since $V_{\rm{eff}}'(R_T)>0$ it follows from the equation above
that $r$ should be negative. Integrating the above equation we get
\begin{equation}
r=-\frac{1}{4} \left(r_0\mp \sqrt{V_{\rm{eff}}'(R_T)}t \right)^2 \
.
\end{equation}
if we demand that for $t=0$ the particle reaches its turning point
we have $r_0=0$ and hence we obtain
\begin{equation}
r=-\frac{V_{\rm{eff}}}{4}t^2
\end{equation}
so that for small negative $t$ (in order to trust $r\ll 1$) the
particle approaches its turning point and then it moves back.

As follows from the properties of the effective potential
$V_{\rm{eff}}$ \footnote{Namely, since the potential reaches zero
for $R\rightarrow 0$ from below and since the potential blows up
for $R\rightarrow \infty$ there should certainly exists the point
where $V'_{\rm eff}(R_m)=0$.} there should also exist the point
where the potential reaches its local minimum corresponding to
$V'_{eff}(R_m)=0$. Again, if we introduce $r$ as $R=r+R_m$  we get
\begin{equation}
V_{\rm{eff}}(R)=V_{\rm{eff}}(R_m)+\frac{1}{2}V''_{\rm eff}
(R_m)r^2
\end{equation}
where $V_{\rm{eff}}(R_m)\equiv A<0$ and $\frac{1}{2}V''_{\rm
eff}(R_m)\equiv B
>0$
and hence the conservation of energy implies the following
differential equation
\begin{eqnarray}
\frac{dr}{\sqrt{1+\frac{B}{A}r^2}}= \pm\sqrt{-A}.
\end{eqnarray}
The above equation has a solution for $r$ that is
\begin{eqnarray}
r=\sqrt{\frac{|A|}{B}}\sin \sqrt{B} t
=\sqrt{\frac{2|V_{\rm{eff}}(R_m)|} {V''_{\rm eff}(R_m)}}\sin
\sqrt{\frac{V''_{\rm eff}(R_m)}{2}}t \ .
\end{eqnarray}
In other words if we have a supertube inserted close to its local
minimum position we observe that the supertube will fluctuate
around this point with harmonic oscillations.

Finally, we will consider the case $R\rightarrow 0$. Since now the
effective potential is given by
\begin{eqnarray}
V_{\rm{eff}}=\frac{R^5}{N}\left( \frac{4\pi^2 p_up^u}
{\mE^2}-1\right) \ , \end{eqnarray} the time evolution equation
for $R$ is
\begin{equation}
\dot{R}^2=-\frac{R^5}{N} \left( \frac{4\pi^2 p_up^u}
{\mE^2}-1\right) \Rightarrow \frac{1}{R^3}=
\left(C\mp\frac{3t}{2}\sqrt{ \frac{1}{N} \left(1-
\frac{4\pi^2p_up^u}{\mE^2}\right)} \right)^2 \ ,
C^2=\frac{1}{R_0^3} \ ,
\end{equation}
whrere $R_0$ is  the position of supertube in time $t=0$. However
since we demand that $R\rightarrow 0$ it is clear that the above
solution is valid in case of large positive or negative $t$ and we
obtain following assymptotic behaviour of $R$
\begin{equation}
R\sim \frac{1}{t^{2/3}} \frac{1}{\left[\frac{9}{4N} \left(1-
\frac{4\pi^2p_up^u}{\mE^2}\right) \right]^{1/3}} \ .
\end{equation}
This result shows that supertube approaches the worldvolume of $N$
D2-branes for $t\rightarrow \infty$.
\subsection{D4-brane background}
The situation of supertube in D4-brane background is similar to
the case of D2-brane background. Namely, the potential $V$ takes
the form
\begin{equation}
V=(R^2+B^2)\left[\frac{\Pi^2R}{R^3+N}+ \tau_2^2\right] + p_up^u
\end{equation}
and hence the effective potential takes the form
\begin{equation}
V_{\rm{eff}}= \frac{R^3}{(N+R^3)} \left(
\frac{4\pi^2}{\mE^2}(R^2+B^2)\left[\frac{\Pi^2R}{R^3+N}+
\tau_2^2\right] + \frac{4\pi^2p_up^u}{\mE^2}-1\right) \ .
\end{equation}
from which we again obtain the asymptotic behavior
\begin{eqnarray}
V_{\rm{eff}}\sim \frac{R^3}{N} \left(\frac{4\pi^2
(B^2\tau_2^2+p_u^2)}{\mE^2}-1\right)
 \ , R\rightarrow 0 \nonumber \\
V_{\rm{eff}}\sim \frac{4\pi^2\tau^2_2}{\mE^2} R \ , R \rightarrow
\infty \ , \nonumber \\
\end{eqnarray}
Since $\frac{\mE^2}{4\pi^2}> (B^2\tau_2^2+p_u^2)$, we see that the
effective potential approaches zero in the limit $R\rightarrow 0$
from below. In the same way we also see that the potential blows
up for $R\rightarrow \infty$. It then follows that the behavior of
the supertube in this background is the same as in the case of
D2-brane background studied in the previous section. Therefore we
skip the details of the discussions in this case.

%%%%%%%%%%%%%%%%%%%%%%%%%%%%%%%%%%%%%%%%%%%%
%%%%%%%%%%%%%%%%%%%%%%%%%%%%%%%%%%%%%%%%%%%%
\subsection{NS5-brane background}
As the next example we will consider supertube in the background
of $N$ NS5-branes.
 The metric, the dilaton and NS-NS field are
\begin{eqnarray}
ds^2 &=& dx_{\mu}dx^{\mu}+H_{NS}(R) dx^mdx^m \ , \cr & \cr
e^{2(\Phi-\Phi_0)}&=& H_{NS}(R) \ , \cr & \cr H_{mnp}&=&
-\epsilon^q_{mnp}\partial_q
\Phi \ , \nonumber \\
\end{eqnarray}
where $H_{NS} = 1 + \frac{N}{R^2}$ is the harmonic function in the
transverse directions of $N$ NS5-branes. If we again restrict to
the homogenous modes the Hamiltonian density takes the form
\begin{eqnarray}
\mH=\sqrt{p_R\frac{1}{H_{NS}} p_R+p_ug^{uv}p_v+\Pi^2+ b_\phi
\frac{1}{R^2H_{NS}}b_\phi+ \frac{\tau_2^2}{H_{NS}}
\left(R^2H_{NS}+B^2\right)} \nonumber \\
=\frac{1}{\sqrt{N+R^2}} \sqrt{p_R^2R^2+p_up^u+(N+R^2+B^2)(\Pi^2+
\tau_2^2R^2)} \ , \nonumber \\
\end{eqnarray}
where we have used
\begin{equation}
b_\phi=F_{\phi z}\pi^z= -B\Pi
\end{equation}
and the fact that  $\pi^z\equiv \Pi$ and $p_u$ are conserved. We
must also stress that we consider the situation when modes $X^m$
that were defined in previous section, are not excited and equal
to zero.

Now the equation of motions take the form
\begin{eqnarray}
\dot{A}_z &=& E=\frac{\sqrt{-g_{00}}}{\sqrt{\mK}}
(g_{zz}+g^{\phi\phi}B^2)\Pi^2 \ , \cr & \cr
\dot{A}_\phi &=& 0 \ , \nonumber \\
\end{eqnarray}
where the second equation above implies that $B=\partial_zA_\phi$
is conserved as well.

In order to study the dynamics of the radial mode we use as in the
previous sections the fact that the energy is conserved. Firstly,
if we write the Hamiltonian density as $\mH=\sqrt{
p_Rg^{RR}p_R+V}$ where
\begin{equation}
V=\frac{1}{N+R^2} \left(p_up^u+ (N+R^2+B^2)
(\Pi^2+\tau_2^2R^2)\right)
\end{equation}
Now we can express $p_R$ from $\mE=2\pi \mH$ as
\begin{equation}
p_R=\frac{\sqrt{g_{RR}}}{2\pi} \sqrt{\mE^2-4\pi^2V}
\end{equation}
and hence the equation of motion for $\dot{R}$ is
\begin{eqnarray}
&&\dot{R}=\frac{p_Rg^{RR}} {\mH}=\frac{1}{\mE\sqrt{g_{RR}}}
\sqrt{\mE^2-4\pi^2 V} \cr & \cr &&\Rightarrow
\dot{R}^2+V_{\rm{eff}}=0 \ ,
\end{eqnarray}
where
\begin{eqnarray}
V_{\rm{eff}}&=& \frac{1}{g_{RR}} \left(\frac{4\pi^2V}{\mE^2}-1
\right)\cr & \cr &=& \frac{R^2}{N+R^2} \left( \frac{4\pi^2}
{(N+R^2)\mE^2} \left(p_up^u+ (N+R^2+B^2)
(\Pi^2+\tau_2^2R^2)\right)-1 \right)
 \ . \nonumber \\
\end{eqnarray}
For small and for large $R$ this potential reduces to
\begin{eqnarray}
V_{\rm{eff}}\sim \frac{R^2}{N} \left(\frac{4\pi^2} {\mE^2 N}
(p_up^u+(N+B^2)\Pi^2)-1\right), R\rightarrow 0 \ , \nonumber \\
V_{\rm{eff}}\sim \frac{4\pi^2 R^2}{\mE^2} \ , R\rightarrow \infty
\ .
\nonumber \\
\end{eqnarray}
Again we see that the potential $V_{\rm{eff}}$ approaches zero for
$R\rightarrow 0$ from below and blows up for $R\rightarrow
\infty$. Consequently we obtain the same picture of the supertube
dynamics as in the previous sections.
%%%%%%%%%%%%%%%%%%%%%%%%%%%%%%%%%%%
\subsection{Macroscopic fundamental string background}
In this section, we study the time dependent dynamics of the tube
in the Macroscopic fundamental string background, where the
metric, dilaton and the NS-NS charge of the F-background is given
by
\begin{eqnarray}
ds^2 = \frac{1}{H_f (r)} \left(-dt^2 + dz^2\right) + \delta_{mn}
dx^m dx^n, \ B_{01} = H^{-1}_f -1, \ e^{2\Phi} = H^{-1}_f,
\end{eqnarray}
where $H_f = 1 + \frac{N}{r^6}$ is the harmonic function in the
transverse eight-space of the $N$ F-strings. $r$ denotes the
spatial coordinate transverse to the macroscopic string.

Let us again consider the Hamiltonian density for D2-brane
\begin{eqnarray}\label{Hdeng}
\mH &=& \sqrt{-g_{00}} \sqrt{\mK}+\Pi^aB_{a0} \ , \cr & \cr \mK&=&
(p_r+\pi^aB_{ar})g^{rs} (p_s+\pi^aB_{as})+ \pi^a\gamma_{ab}\pi^b+
b_ag^{ab}b_b+e^{-2\Phi}\tau_2^2 \det\bA_{ab} \  , \cr & \cr
 b_a&=&\Pi_s\partial_aX^s+ F_{ab}\pi^b \ .
 \nonumber \\
\end{eqnarray}
The Hamiltonian density (\ref{Hdeng}) is the starting point for
our calculation where we consider D2-brane supertube stretched in
$z$ direction and that wind $\phi$ direction in the plane
$(x^8,x^9)$ in the space transverse to the fundamental string,
where the modes that parameterize the embedding of the supertube,
are time dependent only:
\begin{equation}
R=R(t) \ , X^m=X^m(t) \ m=3,\dots, 7  \ .
\end{equation}
Now thanks to the manifest rotation symmetry $SO(6)$ of the
subscpace $(x^3,\dots,x^7)$ we can restrict ourselves to the
motion of supertube in $(x^3,x^4)$ plane where we introduce polar
coordinates as
\begin{equation}
X^3=\rho \cos\psi \ , X^4=\rho \sin\psi \ .
\end{equation}
Note also that the fact that we consider homogenous modes implies
that $A_z$ and $\Pi$ are time independent.

Now the spatial matrix $\bA_{ab}$ takes the form
\begin{equation}
\bA=\left(\begin{array}{cc}
g_{zz} & B_{z\phi} \\
B_{\phi z} & g_{\phi\phi} \\
\end{array}\right)=
\left(\begin{array}{cc}
H_f^{-1} & B \\
-B & R^2 \\
\end{array}\right)
\end{equation}
hence its determinant is equal to
\begin{equation}
\det\bA_{ab}=H_f^{-1}R^2+ B^2  \ ,
H_f=1+\frac{N}{(R^2+\rho^2)^{3}} \ .
\end{equation}
We also consider the electric flux in the $z$ direction only
$\pi^z\equiv \Pi$ so that the term $\pi^aB_{aM}= \pi^zB_{zM}=0$
using the fact that all fields do not depend on $z$. In the same
way $b_\phi=F_{\phi z}\pi^z= B\Pi$. Now the term $\mK$ takes the
form
\begin{eqnarray}
\mK=p_R^2+p_\rho^2+\frac{p_\psi^2}{\rho^2} +\frac{\Pi^2}{H_f}
  +\frac{B^2\Pi^2}{R^2}
+\tau_2^2(R^2+H_fB^2) \ .
\nonumber \\
\end{eqnarray}

It is now easy to see that we can restrict ourselves to the motion
when $p_\rho=p_\psi=\rho=0$. Then we get
\begin{eqnarray}
\mK=\frac{(R^8+B^2N +B^2R^6)}{R^6} \left(\frac{R^6p_R^2}{
R^8+B^2N+B^2R^6}+ \frac{\Pi^2R^4+B+R^6}{N+R^6} \right) \ .
\nonumber \\
\end{eqnarray}

Then finally the Hamiltonian density takes the form
\begin{eqnarray}
\mH&=&
 \frac{1}{\sqrt{R^6+N}} \sqrt{( R^8+B^2N +B^2R^6)
\left(\frac{R^6p_R^2} { R^8+B^2N+B^2R^6}+ \frac{\Pi^2R^4+B+R^6}
{N+R^6} \right)}\cr & \cr &&
\>\>\>\>\>\>\>\>\>\>\>\>\>\>\>\>\>\>\>\>\>\>\>\>\>\>\>\>\>\>\>\>\>\>\>
\>\>\>\>\>\>\>\>\>\>\>\>\>\>\>\>\>\>\>\>\>\> + \frac{\Pi N}
{N+R^6} \ .
\nonumber \\
\end{eqnarray}
In order to simplify notation we will write the
 Hamiltonian density for supertube in Fstring background
 in the form
\begin{equation}
\mH=\sqrt{F(R)p_R^2+G(R)}+ \frac{\Pi N}{R^6+N} \ ,
\end{equation}
where
\begin{equation}
F=\frac{R^6}{R^6+N} \ , G(R)=\frac{(R^8+R^6B^2+B^2N)
(R^6+\Pi^2R^4+N)}{(R^6+N)^2} \ .
\end{equation}
Then the differential equation for $R$ is
\begin{equation}
\dot{R}= \frac{Fp_R}{ \sqrt{(\dots)}} = \frac{Fp_R}{
\left(\frac{\mE}{2\pi}-\frac{\Pi N} {R^6+N}\right)}
\end{equation}
where we have expressed the  square root using the conserved
energy $\mE$. If we also express $p_R$ using $\mE$ as
\begin{equation}
p_R^2=\frac{1}{F} \left[ \left(\frac{\mE}{2\pi}- \frac{\Pi N
F}{R^6}\right)^2-G\right]
\end{equation}
we get
\begin{eqnarray}
\dot{R}^2&=&\frac{F^2p_R^2} {\left(\frac{\mE} {2\pi}-\frac{\Pi N
F} {R^6}\right)^2}= \frac{F}{\left(\frac{\mE} {2\pi}-\frac{\Pi N
F} {R^6}\right)^2}\left[ \left(\frac{\mE} {2\pi}-\frac{\Pi N F}
{R^6}\right)^2-G\right] \Rightarrow \cr & \cr
V_{\rm{eff}}&=&\frac{F}{\left(\frac{\mE} {2\pi}-\frac{\Pi N F}
{R^6}\right)^2}\left[G- \left(\frac{\mE} {2\pi}-\frac{\Pi N F}
{R^6}\right)^2\right] \ . \nonumber \\
\end{eqnarray}
Now we can perform the analysis exactly in the same way as in the
previous section. Namely, the asymptotic behavior of
$V_{\rm{eff}}$ is
\begin{eqnarray}
V_{\rm{eff}}&\sim& \frac{R^6}{N}\frac{1}{
(\frac{\mE}{2\pi}-\Pi)^2} \left[B^2-\left(\frac{\mE}{2\pi}-\Pi
\right)^2\right] \ , R\rightarrow 0 \ , \cr &\cr
V_{\rm{eff}}&\sim& \frac{R^24\pi^2}{\mE^2} \ , \ \ \ \ \ \ \ \ \ \
\ \ \ \ \ \ \ \ \ \ \ \ \ \ \ \ \ \ \ \ \ \ \ \ \
R\rightarrow \infty \nonumber \\
\end{eqnarray}
The potential approaches zero from below for $ R\rightarrow 0$ on
condition
\begin{equation}
\frac{\mE}{2\pi}>B+\Pi
\end{equation}
that is again obeyed. Consequently the analysis is the same as in
the examples studied in the previous sections.

\section{Summary and Conclusion}
In this paper, we have studied the time dependent dynamics of the
supertube in various backgrounds, by using the Dirac-Born-Infeld
effective field theory description in Hamiltonian formalism. This
provide yet another example of time dependent solutions of string
theory in terms of some exotic bound of various D-branes with
Born-Infeld electric and magnetic fields. The stability of these
bound states in various curved backgrounds that are discussed in
the literature recently, has been very suggestive in studying such
dynamics and to view the trajectories of the supertube. We adopt
the Hamiltonian formalism for studying such dynamics of D-brane in
general curved backgrounds in the presence of worldvolume gauge
field. By using the crucial gauge fixing, we have studied the
dynamics in Dp-brane, NS5-brane and fundamental string backgrounds
and have analyzed the effective potential and the trajectory. It
would be very interesting to study some other exotic bound states
of branes (e.g. it has been argued in \cite{Verlinde:2003pv} that
a string network can be blown up into a D3-brane, the so called
supertube) in general curved background. There the analysis in
terms on Hamiltonian formalism seems much more complicated, but
nevertheless a doable problem. It would also be very interesting
to study the dynamics in Anti-de Sitter (AdS) spaces, and to study
the effect in the dual conformal field theory by using the
gauge-gravity duality. In the case of D-branes in AdS backgrounds,
the tachyon like radion would be spatial dependent and hence
studying the dynamics ought to shed light in studying the various
properties of D-brane in curved backgrounds. We hope to return one
or more of these issues in near future.

\vskip .5in

\noindent {\bf Acknowledgements}: We would like to thank M.
Bianchi, U. Lindstrom, Rikard von Unge and A. Sagnotti for various
useful discussions. J.K. would like to thank the Dipartimento di
Fisica, ''Tor Vergata'' for the kind hospitality where a part of
the work was done. The work of J.K. was supported by the Czech
Ministry of Education under Contract No. MSM 0021622409. The work
of K.L.P. was supported in part by INFN, by the MIUR-COFIN
contract 2003-023852, by the EU contracts MRTN-CT-2004-503369 and
MRTN-CT-2004-512194, by the INTAS contract 03-51-6346, and by the
NATO grant PST.CLG.978785.

%%%%%%%%%%%%%%%%%%%%%%%%%%%%%%%%%%%%%%
%%%%%%% Thebibligraphy %%%%%%%%%%%%%%%%%%%%%
%%%%%%%%%%%%%%%%%%%%%%%%%%%%%%%%%%%%%

\end{document}